\documentclass[11pt,a4paper]{article}
\usepackage{jinstpub}
\usepackage{booktabs}
\usepackage{array}
\usepackage{tikz}
\usepackage{graphicx}
\usetikzlibrary{positioning,fit,arrows.meta}
\usetikzlibrary{positioning,calc,arrows.meta}
\usetikzlibrary{arrows.meta,positioning,fit}
\usepackage{placeins}
\usepackage{subcaption}

\notoc

\title{\boldmath A-GHOST: High-rate streaming of trigger-level data to programmable GPU inference}

\author[1]{I. Xiotidis}
\author[1]{, N. Clarke Hall}
\author[2]{, M. S. Larson}
\author[3]{, R. Gurunathan}
\author[3]{, C. Burdick}
\author[1,2]{, T. Du}
\author[4]{, N. Konstantinidis}
\author[6]{, K. Kordas}
\author[5]{, D. Leshchev}
\author[2,5]{, D. W. Miller}
\author[1,4]{, V. A. Petrovic}
\author[7]{, D. Sampsonidou}
\author[3]{, A. Thompson}
\author[1]{, and T. Wengler}

\affiliation{Department of Experimental Physics, CERN, Geneva, Switzerland}
\affiliation{Department of Physics, Enrico Fermi Institute, University of Chicago, Chicago, Illinois, USA}
\affiliation{NVIDIA Corporation, Santa Clara, California, USA}
\affiliation{Department of Physics and Astronomy, University College London (UCL), London, UK}
\affiliation{Kavli Institute for Cosmological Physics, University of Chicago, Chicago, Illinois, USA}
\affiliation{Department of Physics, Aristotle University of Thessaloniki, Thessaloniki, Greece}
\affiliation{Institute of Fundamental Science, University of Oregon, Eugene, Oregon, USA}

\emailAdd{ioannis.xiotidis@cern.ch}

\abstract{A-GHOST (A Global Heterogeneous Online Scouting Trigger) is an R\&D effort investigating high-rate streaming readout architectures for high-energy physics (HEP) experiments. The central idea is to convert the deterministic front-end and fixed-latency processing close to the detector from a decision-making platform to an aggregation and streaming source. Compact trigger-level data are streamed to a GPU-enabled backend, where substantially more complex algorithms can run beyond the resource and latency envelope of the hardware trigger. This paper presents a proof-of-concept backend using the NVIDIA IGX Thor development kit. A software rate-controlled transmitter sends data through a QSFP interface and external loopback cable to a second QSFP interface, allowing the network-to-GPU path to be studied before integration with FPGA sources. NVIDIA DAQIRI enables reception directly into GPU-accessible memory, while a custom CUDA kernel reassembles packet payloads into persistent, contiguous TensorRT input windows without data-type conversion, which is handled by the models. Using a HEP-derived event representation consisting of the ten leading calorimeter clusters, the backend scales from 40 to 100 Gbps while sustaining a constant input rate and an inference latency of 0.118 ms at the p95 interval. The system operates for multiple hours without drops or inference errors. Generative and temporally aware neural networks demonstrate workloads beyond those normally deployed in fixed-latency FPGA trigger systems. The result establishes A-GHOST as a basis for future FPGA-to-GPU streaming readout systems.}


\arxivnumber{1234.56789} 

\newcommand{\pt}{p_\text{T}}
\newcommand{\Et}{E_\text{T}}
\newcommand{\HH}{$HH\rightarrow4b$}
\newcommand{\x}[1]{\mathbf{x}_{#1}}
\newcommand{\mean}{\boldsymbol{\mu}}
\newcommand{\stddev}{\boldsymbol{\sigma}}
\newcommand{\z}{\mathbf{z}}

\begin{document}

\maketitle
\flushbottom

\section{Introduction} \label{sec:Introduction}

Modern HEP experiments must reduce detector data rates by many orders of magnitude before permanent storage. At the Large Hadron Collider (LHC), proton--proton bunch crossings occur at 40 MHz, and the first trigger stage is therefore implemented in custom electronics with deterministic latency and tightly bounded resource usage~\cite{atlas_tdr}. This architecture is highly effective, but the earliest trigger algorithms necessarily operate under severe constraints on memory, arithmetic complexity, state, and execution time. Trigger-level analysis, or data scouting, alleviates part of this limitation by recording compact online representations at rates much higher than those possible for full detector events~\cite{atlas_tla, cms_scouting}.

A-GHOST is an R\&D effort aimed at prototyping a complementary streaming-readout architecture. Rather than replacing the fixed-latency trigger, A-GHOST investigates whether compact data products derived from that environment can be streamed continuously to heterogeneous compute at very high rates. The deterministic hardware remains responsible for timing-critical aggregation and preprocessing, while the GPU backend provides a software-defined stage in which larger neural networks, temporal algorithms and more rapidly evolving reconstruction methods can be studied. In this programmable domain, the backend can exploit substantially larger memory, maintain context across multiple events, trade inference batching latency against throughput, and evolve independently of the fixed-latency trigger firmware. Such workloads are difficult to accommodate inside the strict latency and resource budget of an FPGA-based trigger but become natural once the data are available in accelerator memory.

The paper presents a proof-of-concept implementation of that architecture. A controlled transmitter produces HEP-derived event data at programmable rates and sends them through one high-speed network port of an NVIDIA IGX Thor Development Kit. A second port receives the stream through an external cable so that the full network interface and memory path are exercised while the external FPGA source is still under development. NVIDIA DAQIRI (Data Acquisition for Integrated Real-time Instruments)~\cite{daqiri} provides the high-throughput receive path, and a custom CUDA kernel converts packet-orientated delivery into persistent contiguous windows suitable for Machine Learning (ML) inference. Generative models and neural networks that exploit temporal correlations are then used to demonstrate the type of ML workloads that the streaming stage can support.

The scope of the present study is deliberately architectural. The loopback source isolates the backend from the details of a particular FPGA transmitter and allows throughput, device-memory movement, and inference to be studied under reproducible conditions. The factorisation has been proven critical, as it allowed for in-depth exploration of data formats and decoding of them in GPU during inference to achieve the highest possible rates. 

\section{A-GHOST proof-of-concept architecture} \label{sec:AGHOSTArchitecture}

\subsection{Hardware platform}

The prototype is implemented on the NVIDIA IGX Thor Development Kit and an edge-computing platform based on the IGX T7000 modules~\cite{nvidia_igx, nvidia_thor_overview}. The system resources are summarised in table~\ref{tab:IGXHardware}.

\begin{table}[t]
\centering
\caption{Relevant hardware resources of the IGX Thor Developer Kit used for the A-GHOST prototype. The quoted values are platform capabilities, not measured application performance.}
\label{tab:IGXHardware}
\small
\begin{tabular}{>{\raggedright\arraybackslash}p{0.25\textwidth} >{\raggedright\arraybackslash}p{0.67\textwidth}}
\toprule
Component & Platform capability \\
\midrule
CPU & 14-core Arm Neoverse-V3AE 64-bit CPU, up to 2.6 GHz \\
Integrated GPU & 2,560-core NVIDIA Blackwell GPU with fifth-generation Tensor Cores \\
System memory & 128 GB 256-bit LPDDR5X, 273 GB/s \\
Discrete GPU & NVIDIA RTX PRO 6000 Blackwell Max-Q Workstation Edition; 24,064-core Blackwell GPU \\
Discrete GPU memory & 96 GB GDDR7, 1,792 GB/s \\
Networking & NVIDIA ConnectX-7 with two QSFP112 interfaces, up to 200 GbE per port \\
PCIe fabric & 32 lanes of up to PCIe Gen5 associated with the ConnectX-7 integrated PCIe switch; ConnectX-7 adapters support PCIe Gen5 x16 interfaces~\cite{connectx7} \\
Storage & 1 TB NVMe SSD \\
\bottomrule
\end{tabular}
\end{table}

The ConnectX-7 is particularly relevant to this study because the two QSFP112 ports and the PCIe Gen5 fabric are co-located with the accelerator resources, meaning that the NIC has direct access to the GPU memory via the RDMA protocol. The IGX Thor kit is well suited for prototyping applications with up to 2 x 200 GbE QSFP112 interfaces through the ConnectX-7, with 32 lanes of up to PCIe Gen5 and an integrated PCIe switch. The measurements reported in this study are using the two network ports in an external loopback configuration and exercise offered rates up to 100 Gbps. The available network hardware therefore provides headroom beyond the rates studied in this first demonstrator.

A-GHOST does not rely on the IGX Thor as a unique final deployment target. It is used here because it provides, in one reproducible platform, the high-speed network interface, PCIe connectivity, CUDA-capable GPUs, and sufficient device memory to test a complete receive-to-inference pipeline. The systems question under study is the same for a future experiment-specific implementation, where data arriving into the GPU are originating from deterministic front-end electronics and can be moved to the accelerator memory domain quickly enough that the programmable processing stage remains useful.

\subsection{Streaming path}

\begin{figure}[b]
\centering

\begin{tikzpicture}[
    >=Latex,
    every node/.style={font=\small},
    box/.style={
        draw,
        rounded corners,
        minimum height=9mm,
        align=center,
        inner xsep=5pt,
        inner ysep=4pt
    }
]


\node[
    box,
    text width=42mm
] (source)
{Fixed-latency source\\
\textbf{current:} rate-controlled TX\\
\textbf{future:} FPGA trigger output};

\node[
    box,
    right=7mm of source,
    text width=12mm
] (nic)
{NIC};

\node[
    box,
    right=7mm of nic,
    text width=18mm
] (tx)
{QSFP112\\TX};


\node[
    box,
    below=14mm of tx,
    text width=18mm
] (rx)
{QSFP112\\RX};

\node[
    box,
    left=7mm of rx,
    text width=35mm
] (daqiri)
{ConnectX-7 + DAQIRI\\
packet reception};

\node[
    box,
    left=7mm of daqiri,
    text width=28mm
] (buffers)
{GPU-accessible\\
receive buffers};


\node[
    box,
    below=14mm of buffers,
    text width=34mm
] (pack)
{CUDA span packer\\
contiguous event windows};

\node[
    box,
    right=8mm of pack,
    text width=23mm
] (ml)
{ML\\inference};


\draw[->, thick] (source) -- (nic);
\draw[->, thick] (nic) -- (tx);

\draw[->, thick] (rx) -- (daqiri);
\draw[->, thick] (daqiri) -- (buffers);

\draw[->, thick] (buffers) -- (pack);
\draw[->, thick] (pack) -- (ml);


\node[
    draw,
    dashed,
    fit=(source)(nic)(tx)(rx)(daqiri)(buffers)(pack)(ml),
    inner sep=7pt,
    label={[font=\small]below:
        IGX Thor system}
] (thor) {};


\draw[thick]
    (tx.east)
    -- ++(12mm,0)
    |- node[pos=0.5, right=2mm, align=center, font=\small]
       {External\\loopback\\cable}
    (rx.east);

\end{tikzpicture}

\caption{
A-GHOST proof-of-concept data path. In the present study, a local rate-controlled source inside the IGX Thor sends data through a QSFP112 transmit port and an external loopback cable into the second QSFP112 port. The received stream is delivered through DAQIRI into GPU-accessible buffers which have been tuned around the optimal batch size for throughput, assembled into contiguous inference windows by a CUDA kernel, and passed to
the ML inference stage.
}
\label{fig:architecture}
\end{figure}
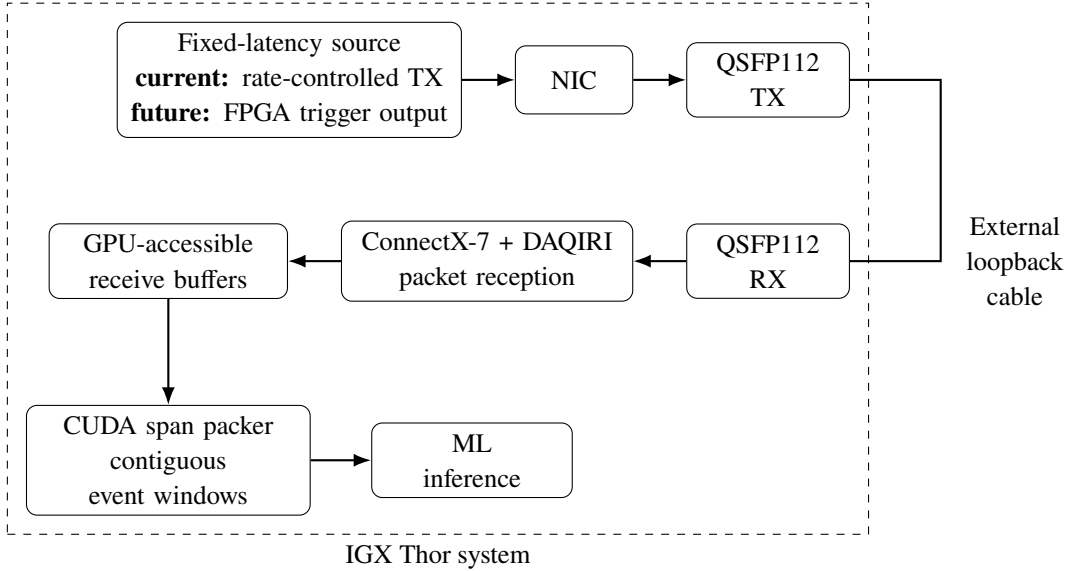

Figure~\ref{fig:architecture} summarises the current data path. A software rate-controlled transmitter, implemented in the IGX Thor hardware, generates the trigger-like event representation consumed by the inference pipeline. The stream leaves one QSFP port, traverses a physical cable, and returns through the second QSFP port. The receive side is implemented with DAQIRI using the ibverbs-backend path employed for the measurements in this paper. This receive path uses GPUDirect RDMA to place incoming payloads directly into GPU-accessible memory while bypassing the conventional host networking stack. Packet reception is therefore performed in user space without kernel-mediated data copies, avoiding an intermediate CPU staging path between the NIC and the accelerator. DAQIRI supports packet reception in user space, bypassing the Linux kernel network stack, as well as GPUDirect memory access for zero-copy NIC-to-GPU data transfer. Finally, ML inference is executed using an optimized NVIDIA TensorRT engine.

The rate control of the transmitter is important for two reasons. First, it allows the networking and inference stages to be tested at well-defined offered rates without coupling the measurement to a specific FPGA Ethernet kernel before the computing part has been deeply understood. Second, it creates a direct path to later hardware integration, as the network-facing interface of the backend does not need to change when the software transmitter is replaced by a deterministic FPGA source.

\subsection{GPU-resident event assembly}

Network reception and neural-network inference operate naturally at different granularities. DAQIRI delivers packet payloads as spans of received data, whereas deployed models are most efficient when presented with larger contiguous input tensors of fixed size. Performing a host copy, allocation, or inference launch for every network fragment would introduce overhead on microsecond timescales and would make the processing rate strongly dependent on packetisation rather than on the model itself. 

To bridge these granularities, our prototype includes a CUDA packing kernel. Packet payloads received into GPU-accessible buffers are assembled directly into contiguous windows for the ML model. The host records source gets into DAQIRI delivery order, while the developed CUDA kernel assigns one thread block to each span and performs a strided device-to-device copy to its precomputed tensor offset. The output events synchronise packing with inference, allowing input assembly and ML execution to overlap without staging packet payloads through host memory. The resulting producer--consumer path decouples network-fragment delivery from the batch size used by the model and avoids allocating or releasing GPU memory on the timescale of individual packets.

This assembly stage is also what enables temporal models. Because the host records spans in delivery order and the GPU writes them into precomputed offsets, consecutive events can be placed into ordered multi-event windows before inference. The same mechanism can therefore support event-wise processing, large batches of independent events, or short temporal contexts without changing the network-facing interface.

Expanding this concept and the operation of the developed CUDA kernel for a multi-source system would require additional development, as potential networking logic can violate the assumptions of contiguous events. The exact addressing of a multi-source system is beyond the scope of the current paper.

\section{HEP-derived data stream}

The study uses simulated proton--proton collisions at $\sqrt{s} = 14$ TeV. Events are generated with \textsc{Pythia}~8.3~\cite{pythia}, and the detector response is simulated with \textsc{Delphes}~\cite{delphes} using a configuration inspired by an ATLAS-like detector at the LHC. A QCD-only sample provides the nominal stream, while gluon-fusion \HH\ events provide a physically motivated anomalous benchmark. The generated sample emulating an event stream including rare physics phenomena, contains 497,500 QCD events and 2,500 \HH\ events. Additional low-energy QCD collisions are overlaid on each event to emulate the expected running conditions at high-luminosity LHC running~\cite{ZurbanoFernandez:2020cco}, with the number of overlaid collisions $\mu$ sampled at random as $\mu\sim\text{Pois}(200)$.
Full details of the datasets can be found in Ref.~\cite{triggeropt}.

The detector response is reduced to a compact trigger-level representation. 
Calorimeter deposits in $\eta \times \phi = 0.1 \times 0.1$ regions
\footnote{We use the LHC-standard $(\pt, \eta, \phi)$ coordinate system, where $\pt$ is the momentum transverse to the beam line, the $\eta$ is the pseudorapidity (linked to the polar angle $\theta$ via $\eta = -\log [\tan (\theta/2)]$), and $\phi$ is the azimuthal angle.} 
with transverse energy $\Et>2$~GeV are clustered with the anti-$k_t$ algorithm~\cite{Cacciari:2008gp}, using $R=0.4$. Clusters are required to have a $\Et > 10$ GeV and are ordered by decreasing $\pt$; the ten highest-$\pt$ clusters are retained. Each event at time step $t$ is therefore represented by
\begin{equation}
    \x{t}=\big\{(\pt,\eta,\phi)_i\big\}_{i=1}^{10},
\end{equation}
corresponding to 30 float32 values, or 120 bytes per event.

This representation is intentionally simple. It is not meant to reproduce the full content of a particular experiment's hardware trigger. Instead, it provides a realistic example of a compact, partially reconstructed physics object stream that could be produced by fixed-latency electronics and exported continuously. It also keeps the systems study interpretable when looking at network rate, event rate, model complexity, and batching. These values can be varied without the confounding effects of a much larger detector payload and also allow the development of potential FPGA source system firmware for optimal throughput. 

\section{ML algorithms}

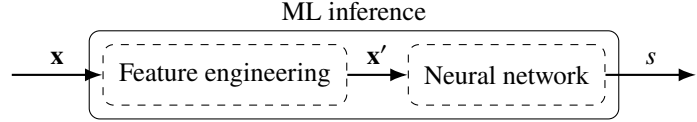
\begin{figure}[h!]
\centering
\begin{tikzpicture}[
    node distance=8mm,
    >=Latex,
    every node/.style={font=\small},
    box/.style={
        draw,
        dashed,
        rounded corners,
        minimum height=8mm,
        align=center,
        inner xsep=6pt
    }
]
\node[box] (features) {Feature engineering};
\node[box,right=of features] (network) {Neural network};

\draw[->,thick]
    ([xshift=-12mm]features.west)
    -- node[midway,above] {$\mathbf{x}$}
    (features.west);

\draw[->,thick]
    (features.east)
    -- node[midway,above] {$\mathbf{x}'$}
    (network.west);

\draw[->,thick]
    (network.east)
    -- node[midway,above] {$s$}
    ([xshift=12mm]network.east);

\node[
    draw,
    rounded corners,
    fit=(features)(network),
    inner sep=5pt,
    label={[font=\small]above:ML inference}
] {};
\end{tikzpicture}
\caption{Detailed diagram of the ML inference step in Figure~\ref{fig:architecture}, showing the sequential feature engineering and neural network pipeline.}
\label{fig:ml_pipeline}
\end{figure}

As the paper aims to demonstrate the technology of streaming readout with high-complexity computing backends, the ML models developed and used in this study are not intended to define the final physics selection for A-GHOST. They are representative online workloads chosen to demonstrate what becomes possible after the streamed data have left the fixed-latency hardware domain. The workload consists of a two-stage pipeline containing feature engineering and neural network stages, as shown in Figure~\ref{fig:ml_pipeline}.

Two workloads are considered, each with a complementary network architecture. The first is an event-wise dense autoencoder (AE). Autoencoders have been evaluated for real-time anomaly detection at the LHC on several occasions~\cite{Govorkova_2022, gelato, axol1tl, cicada, artemis}. Whereas the complexity of these models is heavily constrained by the fixed-latency hardware environment, our model can exceed the parameter count and arithmetic intensity constraints in order to maximise discrimination power, incorporating additional feature engineering and implementing the full reconstruction loss as an anomaly score via a decoder. 

The second is a temporal convolutional neural network (tCNN). The temporal model is a variational autoencoder (VAE)~\cite{kingma2022} that selects for anomalies within three consecutive time-ordered events. The temportal VAE demonstrates two additional key capabilities of the A-GHOST system that exceed those of the fixed latency domain. In employing a VAE, we demonstrate both the sampling and inference stages of a large and non-deterministic generative ML model at the luminosity frontier. The full evaluation of these generative models, including latent-space sampling and decoding, lies beyond the latency and resource budgets of the FPGA-bound trigger system. Previous implementations of such architectures on FGPAs have resorted to model truncation~\cite{Govorkova_2022, gelato, axol1tl, artemis, Vaselli_2026} or knowledge distillation~\cite{cicada, tahseen2026distillingnormalizingflowsrealtime} in order to operate within the fixed latency domain, with both approaches degrading the discrimination power of the parent algorithm. Furthermore, the aggregation of consecutive events allows for the identification and characterisation of time-dependent effects that would otherwise be invisible within a fixed latency time-multiplexed environment. Such a capability could be transformational for both the physics reach and control mechanism of HEP experiments.

\subsection{Feature engineering}
The features are engineered to remove boundaries and Gaussianise the inputs to the neural network. The azimuthal coordinates are rotated according to $\phi\rightarrow\Delta\phi=\mathrm{wrap}_{[-\pi,\pi)}(\phi-\phi_{\Sigma})$, where $\phi_{\Sigma}$ is the azimuth of the vector sum of the cluster momenta. The engineered feature vector $\x{t}'$ is then the result of the following transformation:

\begin{equation}
    \x{t}\rightarrow\x{t}'=\left\{\left(\log(\pt-10),\tanh^{-1}(\eta/5),\tanh^{-1}(\Delta\phi/\pi)\right)_i\right\}_{i=1}^{10}.
\end{equation}
Which operates on jets with $\pt > 10$ GeV. The resulting 30-dimensional vector is standardised to zero mean and unit variance. 

\subsection{Dense event-wise autoencoder}
\begin{figure}[h]
    \centering
    \includegraphics[width=0.85\textwidth]{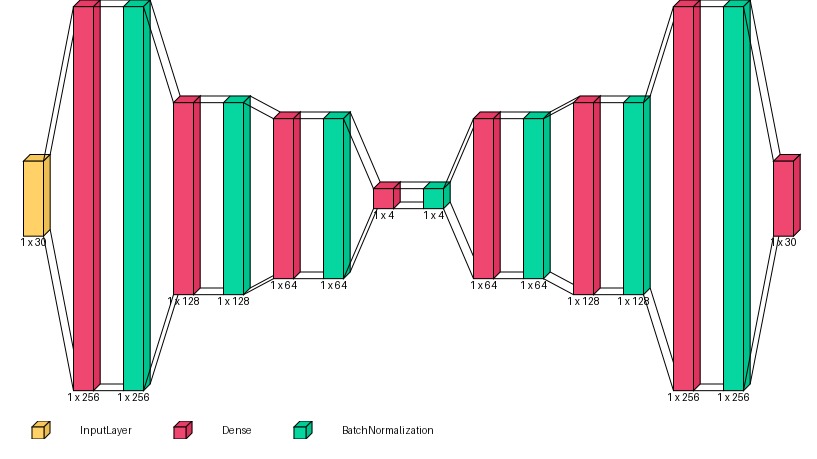}
    \caption{The architecture of the dense event-wise autoencoder used as the first representative GPU workload. A 30-component event vector is mapped through 256-, 128- and 64-unit dense layers to a 4-dimensional latent space and reconstructed symmetrically.}
    \label{fig:ae_model}
\end{figure}

The dense autoencoder neural network (DNN+AE) compresses the event through a series of dense layers with dimensions $256\to128\to64\to4$ and mirrors the structure in the decoder back to a 30-component output. Batch normalisation and the ReLU activation are used throughout. A block diagram of the model is shown in Figure~\ref{fig:ae_model}. The architecture contains approximately 101k trainable parameters and requires 97.8~kMAC per event. The score used to select anomalous events is the mean-squared reconstruction error on $\x{t}'$:
\begin{equation}
    s_t^\text{AE} = \frac{1}{30}|| \x{t}' - \text{AE}(\x{t}')||^2
    \label{eqn:recon_error}
\end{equation}

\begin{figure}[h]
    \centering
    \includegraphics[width=0.6\textwidth]{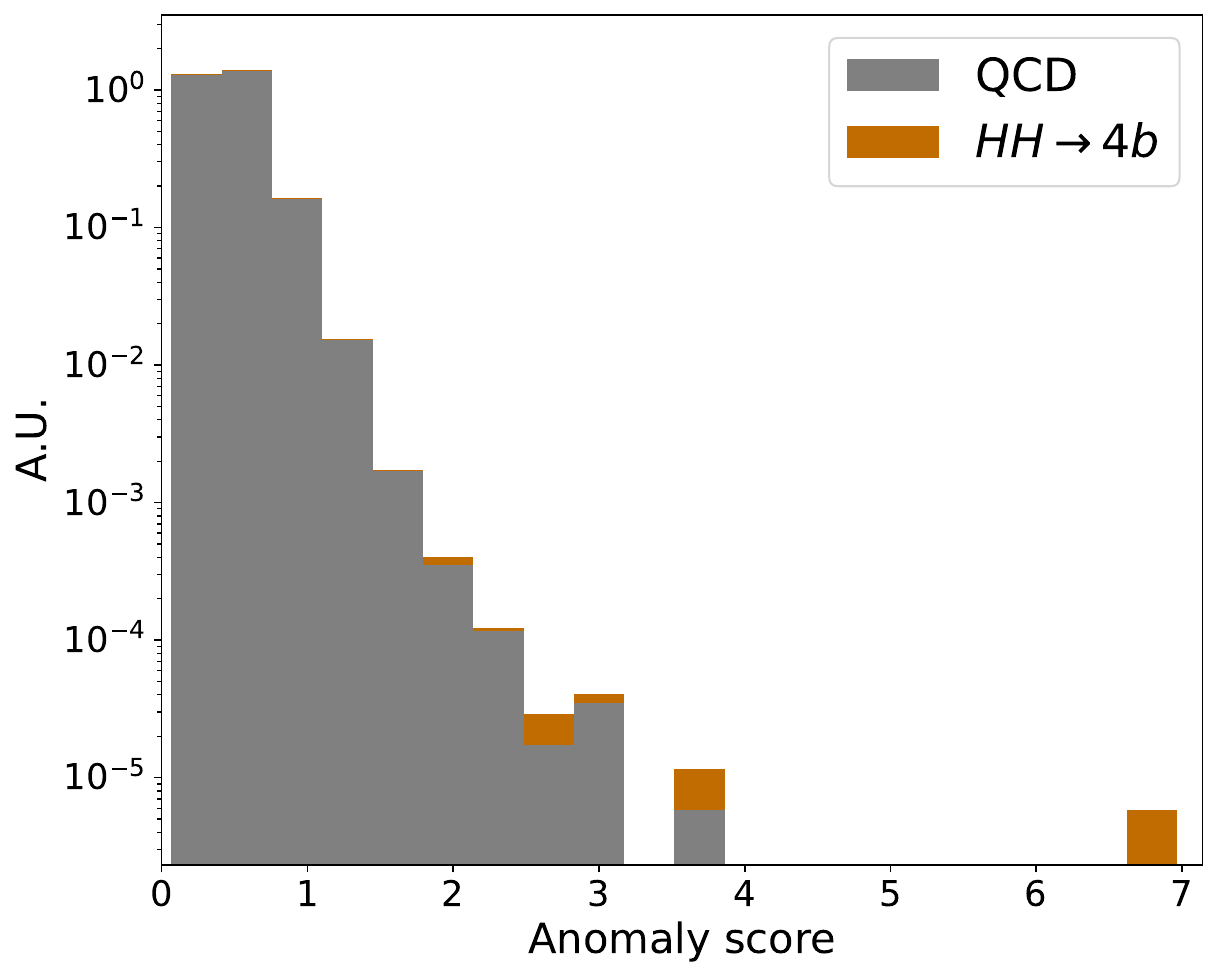}
    \caption{Anomaly-score distribution for the dense autoencoder workload. The nominal QCD sample is shown together with the $HH\to4b$ benchmark.}
    \label{fig:densescore}
\end{figure}

Figure~\ref{fig:densescore} shows the anomaly-score distribution for the nominal QCD sample and the $HH\to4b$ benchmark. The plot is included here as a functional validation of the physics data path rather than as an optimised anomaly-detection result: after packetisation, reception, GPU assembly and inference, the reconstructed score retains non-trivial sample-dependent structure. 

\subsection{Temporal variational autoencoder}

\begin{figure}[h]
    \centering
    \includegraphics[width=0.85\linewidth]{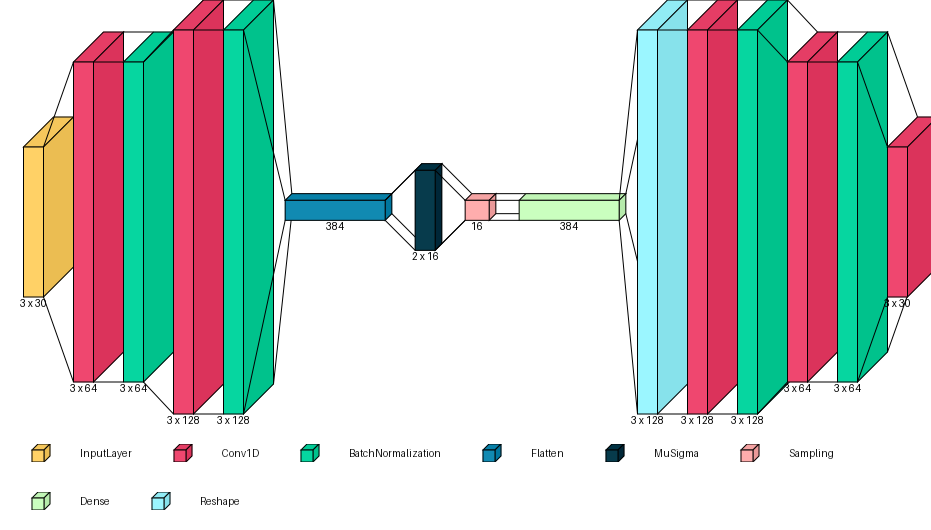}
    \caption{The architecture of the temporal CNN model used to exercise ordered multi-event inference. The network receives a $3\times30$ window and reconstructs the original event sequence.}
    \label{fig:temporalarch}
\end{figure}

The temporal variational autoencoder (tCNN+VAE), depicted in Figure \ref{fig:temporalarch}, operates on three consecutive 30-component event vectors $\x{t-1:t+1}'$, explicitly introducing information across neighbouring bunch crossings. Time-axis convolution and batch normalisation stages expand the representation to 64 and then 128 channels. The resulting $3\times128$ representation is flattened to a 384-component vector, on which two dense layers predict the mean $\mean$ and standard deviation $\stddev$ of a 16-dimensional Gaussian latent distribution. The decoder samples a point $\z\sim N(\mean, \stddev)$ from this Gaussian latent distribution, which is then re-expanded to restore the temporal structure and reconstructed back to the three 30-component event vectors. The score used to select anomalous three-event sequences combines the reconstruction error defined in Equation~\ref{eqn:recon_error} with a Kullback–Leibler (KL) divergence term \cite{Kullback1951}  

\begin{equation}
\label{eqn:vaescore}
    s_{t-1:t+1}^{\text{VAE}} = s_{t-1:t+1}^{\text{AE}} + \beta\times D_{\text{KL}},
\end{equation}
where $\beta$ is a scalar that weights the KL term; here $\beta$ is chosen as 0.01. The KL divergence term, which regularises the latent distribution of normal events towards a unit Gaussian distribution, is defined in terms of $\mean$ and $\stddev$ as
\begin{equation}
  D_{\text{KL}}
  \;=\;
  \frac{1}{2}
    \Bigl(\mean^{2}
    + \stddev^{2}
    - \ln\stddev^{2}
    - 1\Bigr).
  \label{eq:kl}
\end{equation}

This workload illustrates a qualitative difference from the DNN+AE. The backend must preserve event ordering, form overlapping or non-overlapping three-event windows, and hold sufficient state to assemble the next input tensor while the current one is being evaluated. Such operations are possible in FPGAs in principle, but they consume buffering, routing, and deterministic-latency budget. Additionally, due to the difficulty of random Gaussian sampling, VAEs in particular are challenging to implement on FPGAs. In A-GHOST they are moved into the programmable accelerator domain, where the available memory is orders of magnitude larger and the algorithm can be changed without redesigning the front-end firmware.

\begin{figure}[h]
    \centering
    \begin{subfigure}[t]{0.49\textwidth}
        \centering
        \includegraphics[width=\linewidth]{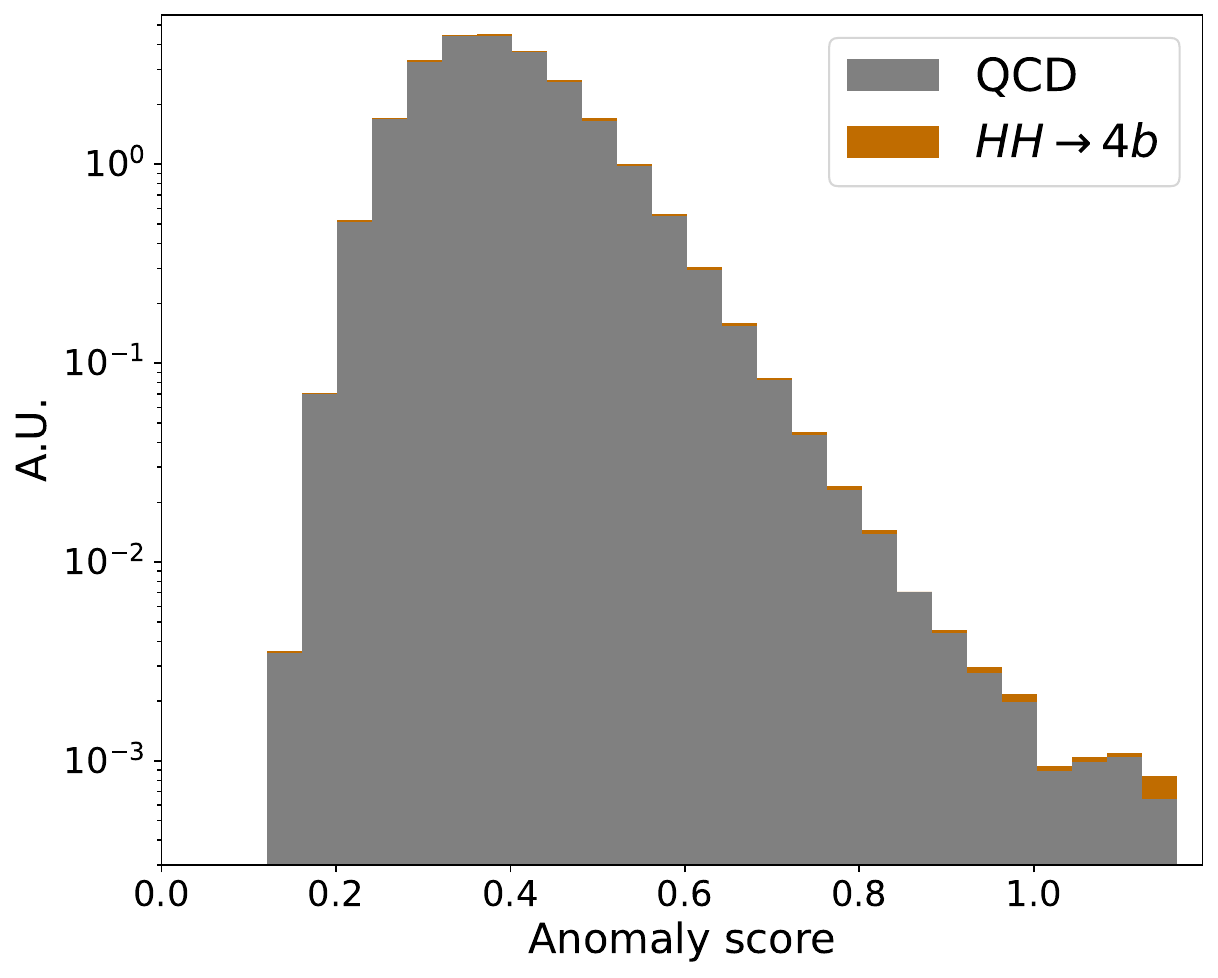}\\[-1mm]
        \caption{tCNN+AE}
        {\label{subfig:tCNN+AE}}
    \end{subfigure}\hfill
    \begin{subfigure}[t]{0.49\textwidth}
        \centering
        \includegraphics[width=\linewidth]{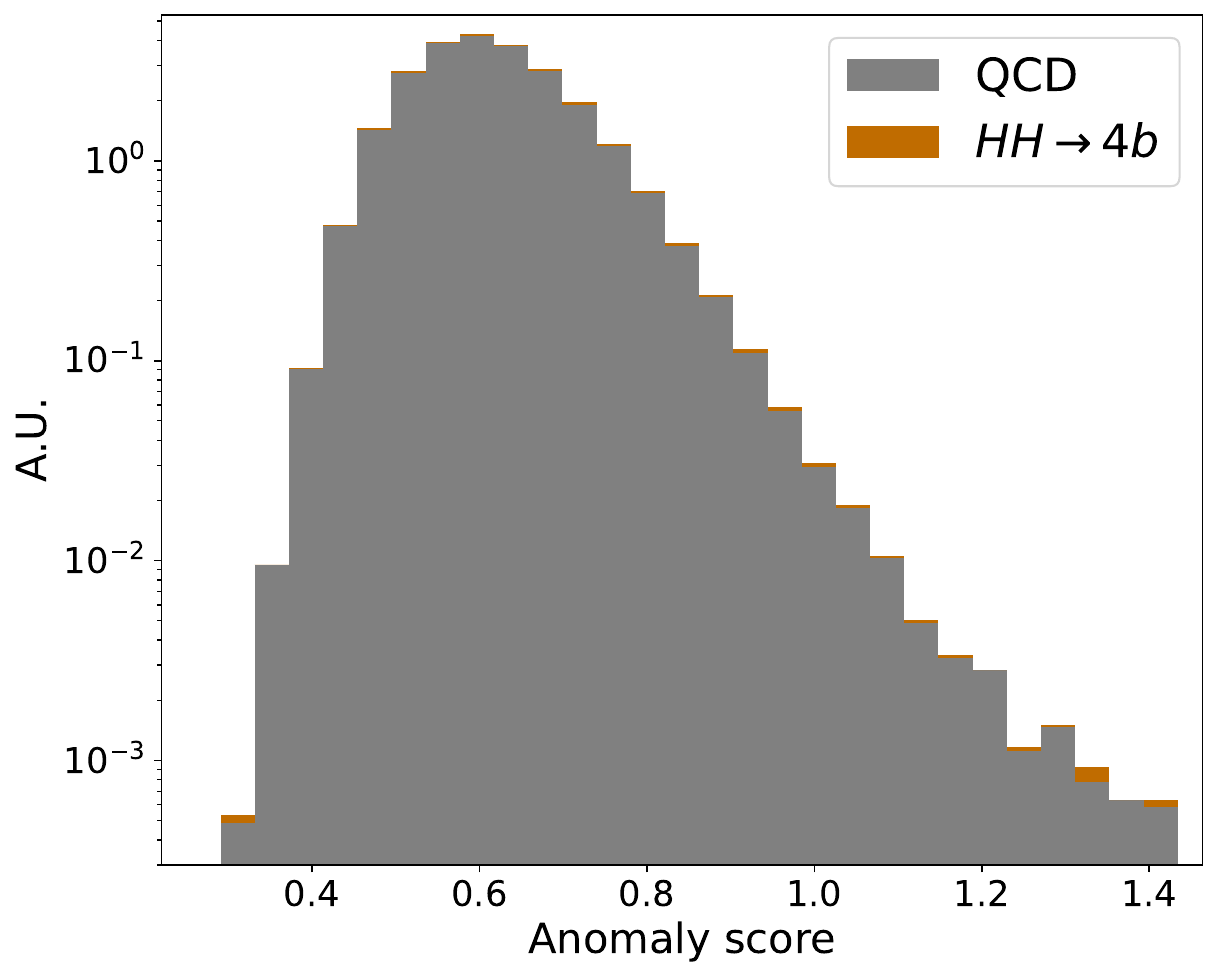}\\[-1mm]
        \caption{tCNN+VAE}
        {\label{subfig:tCNN+VAE}}
    \end{subfigure}
    \caption{Anomaly-score distributions for the temporal CNN workloads. Figure~\ref{subfig:tCNN+AE} shows the reconstruction error from the autoencoder, whereas Figure~\ref{subfig:tCNN+VAE} depicts the full loss of the generative variational autoencoder. The nominal QCD sample is shown together with the $HH\to4b$ benchmark.}
    \label{fig:temporalresult}
\end{figure}

A tCNN + standard autoencoder (tCNN+AE) is also considered. The tCNN+AE shares the encoder-decoder architecture depicted in Figure \ref{fig:temporalarch}; however, it differs in that the tCNN+AE encodes each input to a single 16-dimensional deterministic latent coordinate. The KL divergence term is not evaluated, with the tCNN+AE using only the reconstruction error defined in Equation~\ref{eqn:recon_error} as the anomaly score. The tCNN+AE has approximately 73k trainable parameters and requires 153.9 kMACs per event, while the tCNN+VAE has approximately 75k trainable parameters and requires 154.9 kMACs per event. Figure \ref{fig:temporalresult} shows the anomaly score distributions for the nominal QCD sample and the $HH\to4b$ benchmark for the tCNN+AE and tCNN+VAE models. 

\subsection{Batched inference characteristics}

\begin{figure}[h]
    \centering
    \begin{minipage}[t]{0.49\textwidth}
        \centering
        \includegraphics[width=\linewidth]{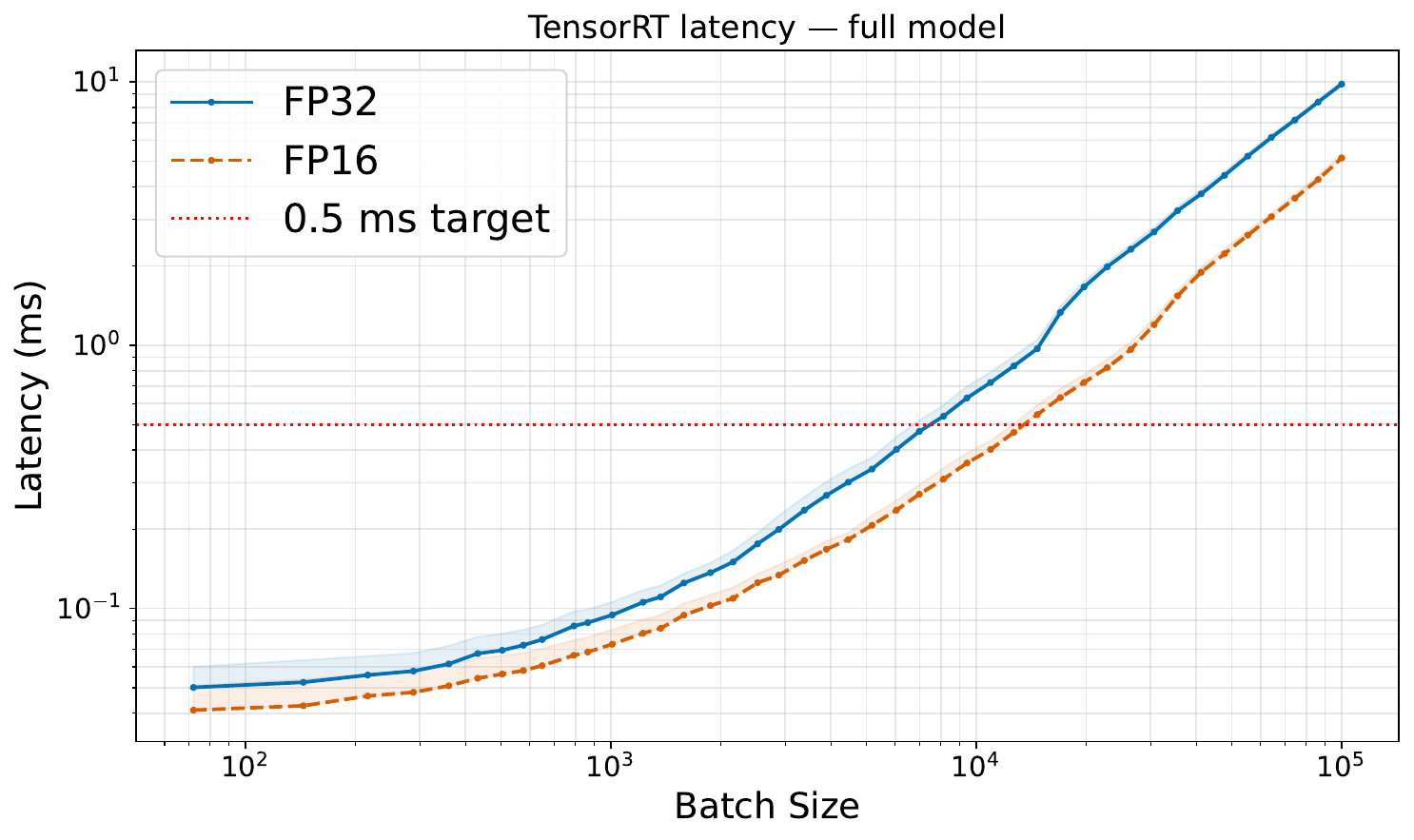}\\[-1mm]
        {\small (a) TensorRT latency.}
    \end{minipage}\hfill
    \begin{minipage}[t]{0.49\textwidth}
        \centering
        \includegraphics[width=\linewidth]{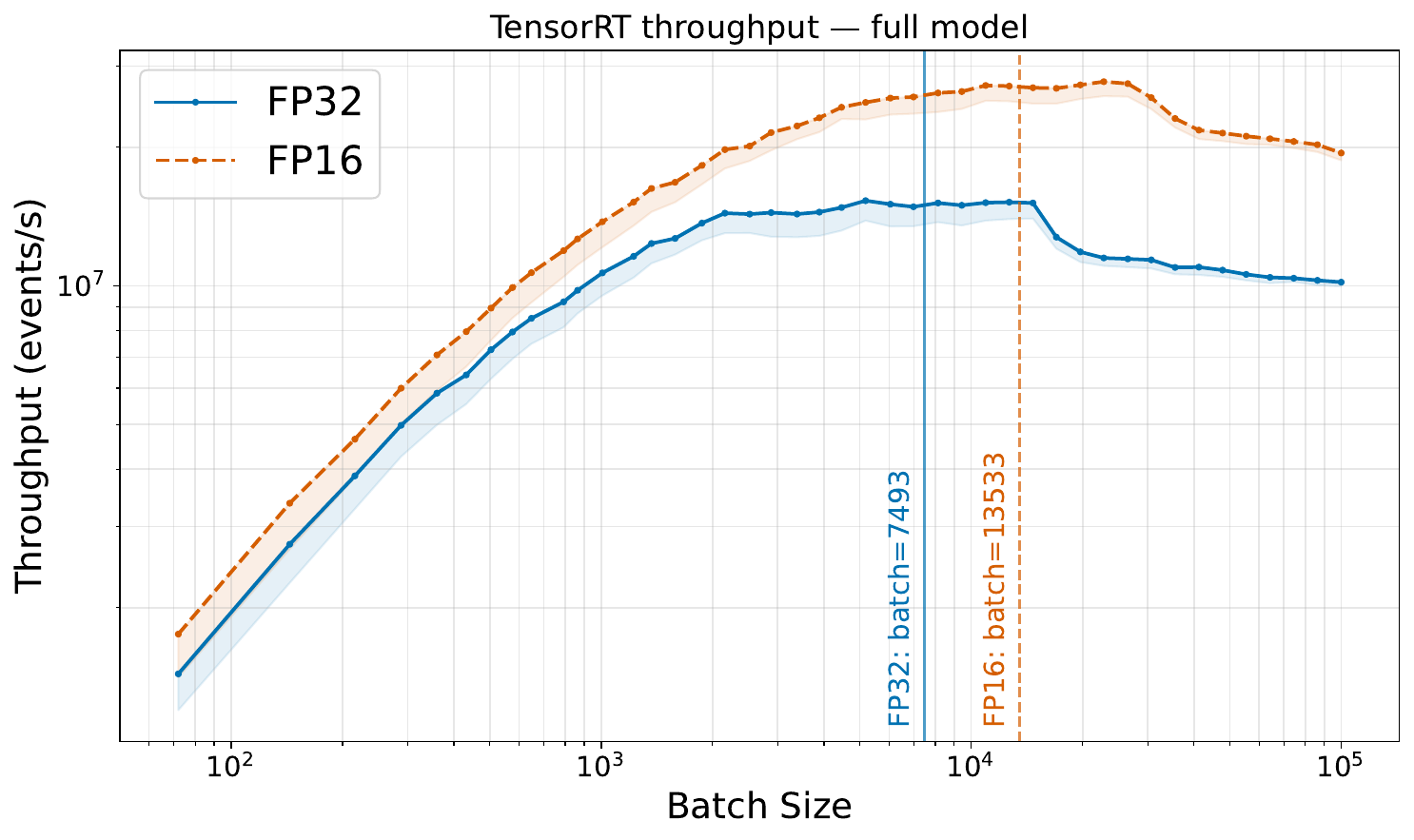}\\[-1mm]
        {\small (b) Event throughput.}
    \end{minipage}
    \caption{Batch size scan for the full temporal workload in FP32 and FP16. The latency panel includes a 0.5 ms reference line. The throughput panel marks representative batches of 7,493 events for FP32 and 13,533 events for FP16. The shaded region above latency and below throughput indicates the p99 interval. }
    \label{fig:batchscan}
\end{figure}

The CUDA assembly stage makes the network delivery granularity independent of the inference batch size. Importantly, batching is confined to the accelerator-side inference stage and does not imply batch formation at the source. In the intended FPGA implementation, events are streamed continuously as they become available. The source-side latency is therefore limited to packetisation/serialisation time associated with the event payload, rather than the time required to accumulate an inference batch. This allows the operating point of the GPU to be selected according to the latency budget of the online application. Figure~\ref{fig:batchscan} shows the measured latency from TensorRT and event throughput of the full temporal workload at float32 and float16 precision as functions of the batch size.

At small batch sizes, inference latency is low but the accelerator is underutilised. In addition, being able to actually release the memory after the execution completes causes the network to stall longer than the delivery and compute time and hence creates packet drops. Increasing the batch size amortises launch and scheduling overheads and raises the effective event throughput. At larger batches, throughput approaches a broad saturation region while the latency of each batch continues to increase. Switching to float16 shifts this trade-off towards higher throughput and remains below the 0.5 ms reference line for a larger batch than float32. The important system-level result is not a single preferred batch size, but rather that the streamed backend can choose this operating point in software, independently of the fixed schedule of the front-end trigger.

\section{Streaming and end-to-end operation}

\subsection{Rate scan}

The rate-controlled source is used to exercise the complete receive and inference pipeline over an offered-rate range of 40--100 Gbps. In the current loopback configuration the receive throughput tracks the requested rate and approaches 100 Gbps at the highest operating point. The measured receive throughput is cross-checked against physical NIC counters to ensure that these packets are entering through the NIC and are placed into GPU-resident input buffers via the GPUDirect path. Host processing remains small compared with the delivered data rate, and no receive drops or inference errors were observed in the test operating points.

\begin{figure}[h]
    \centering
    \includegraphics[width=0.72\textwidth]{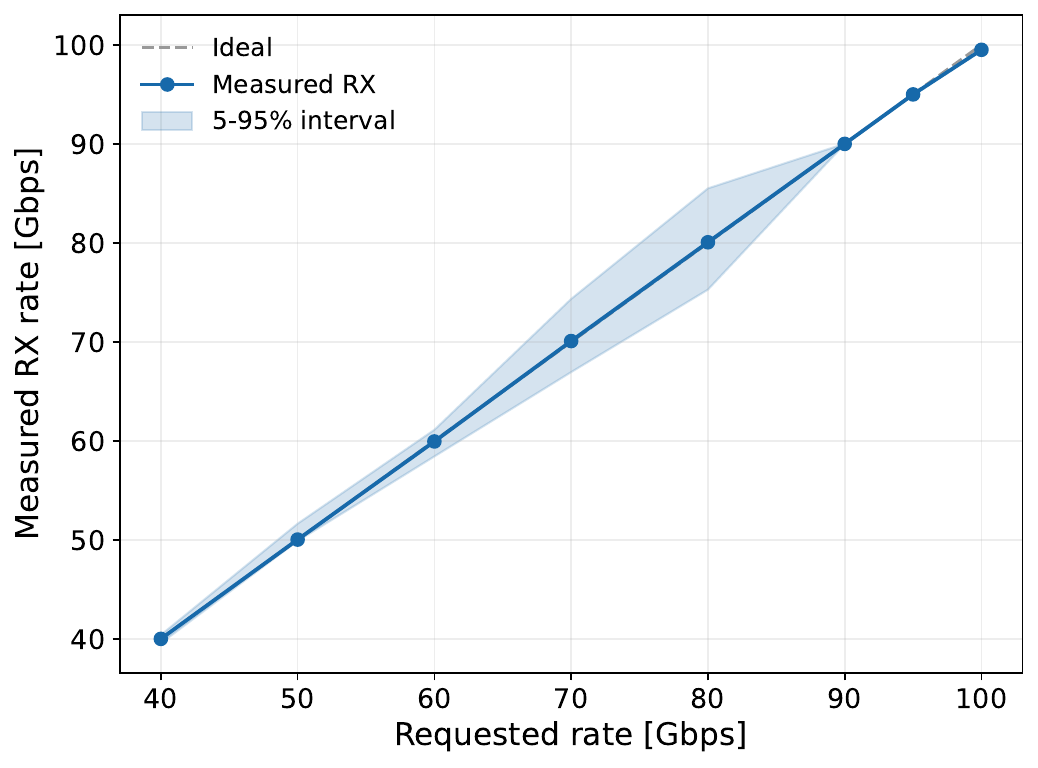}
    \caption{Measured receive throughput as a function of the requested transmit rate for the A-GHOST loopback configuration. 
    The input rate is scanned from 40 to 100 Gbps using the rate-controlled transmitter. 
    The solid line shows the mean received throughput, the dashed line indicates ideal one-to-one scaling, and the shaded band denotes the 5–95\% interval of the steady-state measurements. 
    The receive rate follows the requested load across the full scan and approaches the nominal 100 Gbps link rate at the highest operating point.}
    \label{fig:ratescan}
\end{figure}

The rate scan seen in Figure~\ref{fig:ratescan} should be interpreted as a backend capacity measurement rather than as an experiment-level trigger benchmark. The transmitter and receiver are on the same IGX Thor and are joined externally, so the result establishes the ability of the software, NIC, PCIe and GPU path to sustain the required traffic before the source is replaced by independent FPGA hardware. The measurements have been extracted via the NIC counters directly given that a physical link is used.

\subsection{Concurrent network, memory and inference activity}

\begin{figure}[h]
    \centering
    \includegraphics[width=0.98\textwidth]{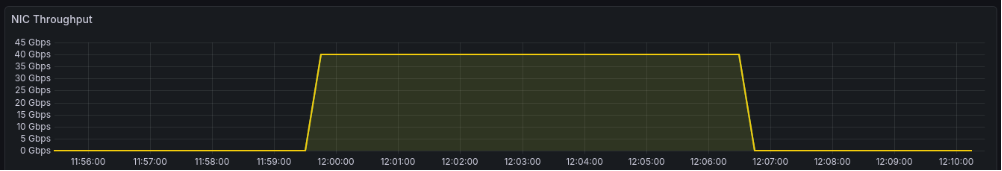}\\[-1mm]
    {\small (a) NIC throughput.}\\[2mm]
    \begin{minipage}[t]{0.49\textwidth}
        \centering
        \includegraphics[width=\linewidth]{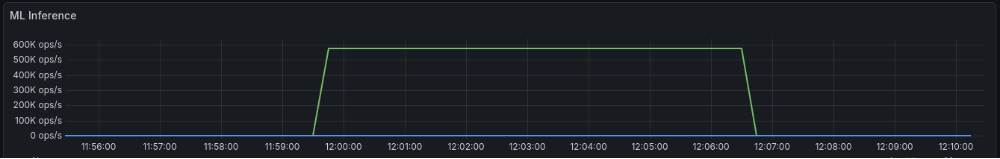}\\[-1mm]
        {\small (b) Inference operation rate.}
    \end{minipage}\hfill
    \begin{minipage}[t]{0.49\textwidth}
        \centering
        \includegraphics[width=\linewidth]{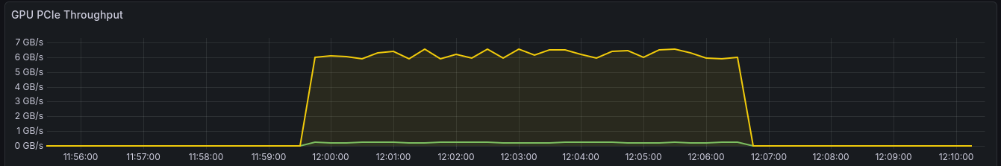}\\[-1mm]
        {\small (c) GPU PCIe throughput.}
    \end{minipage}
    \caption{Live monitoring during a representative sustained 40 Gbps run. The traces show simultaneous network input, inference activity and GPU-side PCIe traffic. The remaining traces in (b) are traces focusing on ML score to storage which wasn't enabled in this run. For (c) the green curve indicates the RX path of the PCIe which only outputs single numbers and hence shows a smaller occupancy of the PCIe bus.}
    \label{fig:live40}
\end{figure}

A representative 40 Gbps run is shown in Figure~\ref{fig:live40}. The network trace remains stable near the requested line rate for the duration of the run. At the same time, the inference service reports a stable operation rate, and the GPU PCIe trace shows sustained device traffic. To be able to extract online information with such a system, a dedicated OpenTelemetry~\cite{opentelemetry} backend has been implemented into DAQIRI along with specific software to expose these metrics to a Grafana~\cite{grafana} monitoring application, called SPOOK. These monitoring plots are diagnostic rather than precision micro benchmarks, but they demonstrate that reception, accelerator-side data movement and model execution remain active concurrently over an extended interval. 

\subsection{100 Gbps stress test}

The 40 Gbps measurement in the preceding subsection is representative of the input conditions considered for the HEP-derived stream used in this study. We separately pushed the backend to an offered rate of 100 Gbps in order to determine the upper end of the present single-link configuration. This measurement is therefore treated as a systems stress test rather than as a realistic experiment-level operating point or a physics-throughput requirement.

\begin{figure}[h]
    \centering
    \begin{minipage}[t]{0.49\textwidth}
        \centering
        \includegraphics[width=\linewidth]{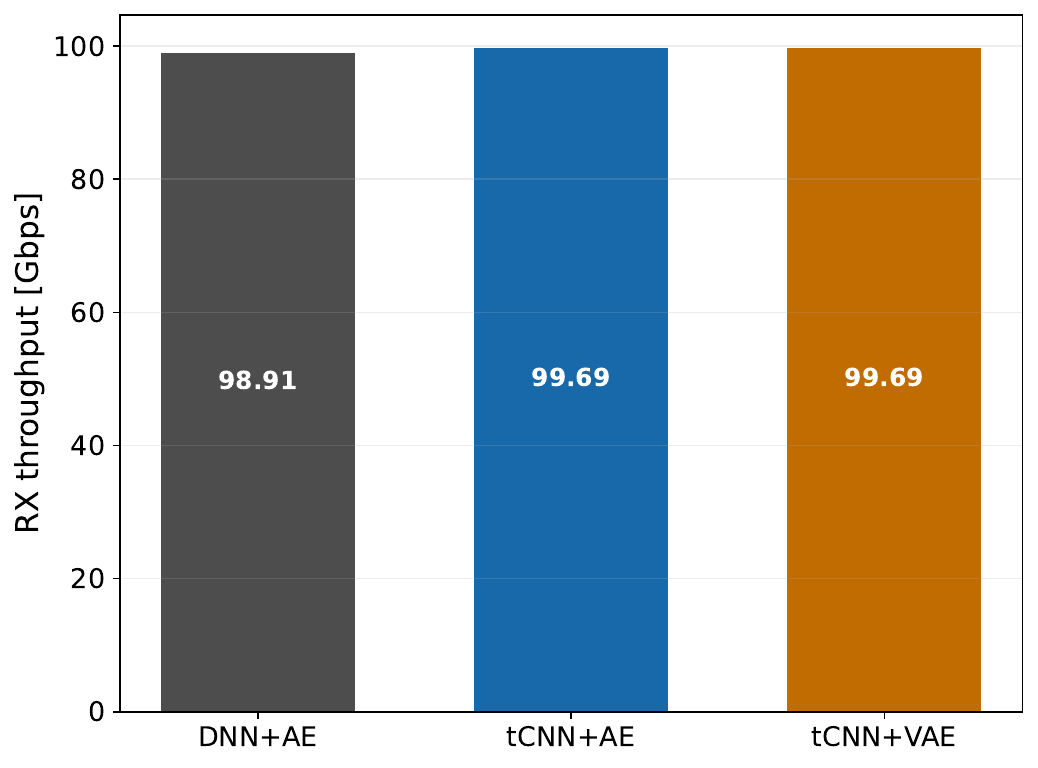}\\[-1mm]
        {\small (a) Delivered receive throughput.}
    \end{minipage}\hfill
    \begin{minipage}[t]{0.49\textwidth}
        \centering
        \includegraphics[width=\linewidth]{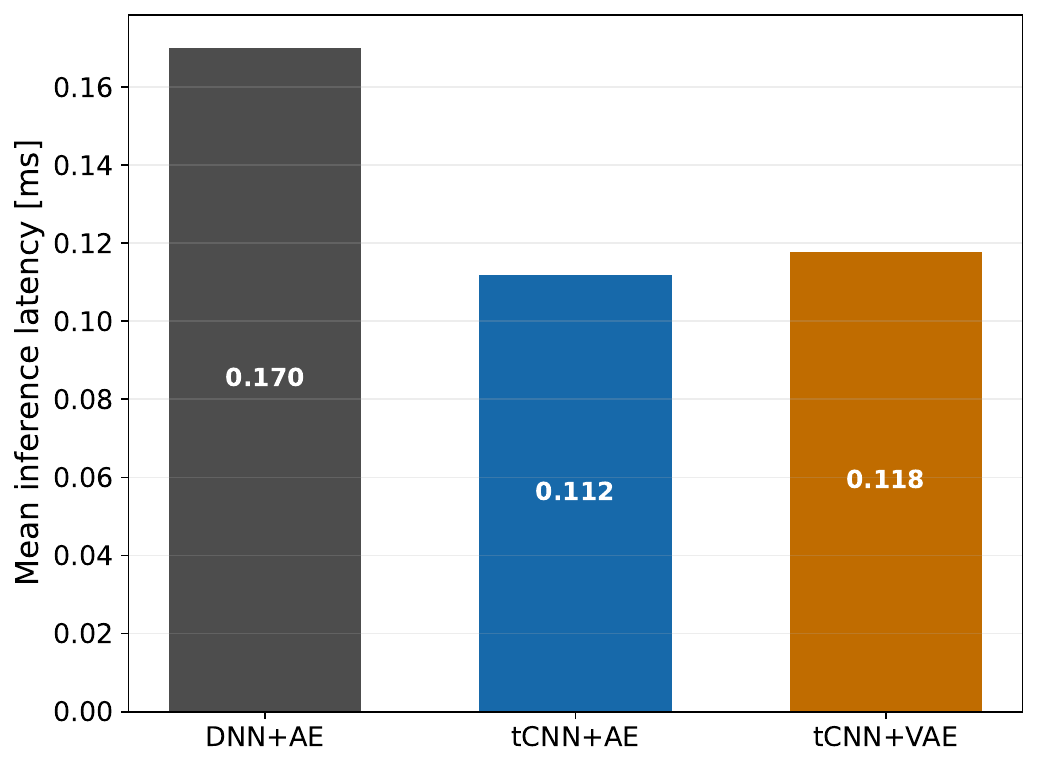}\\[-1mm]
        {\small (b) Mean TensorRT inference latency.}
    \end{minipage}

    \vspace{2mm}
    \begin{minipage}[t]{0.49\textwidth}
        \centering
        \includegraphics[width=\linewidth]{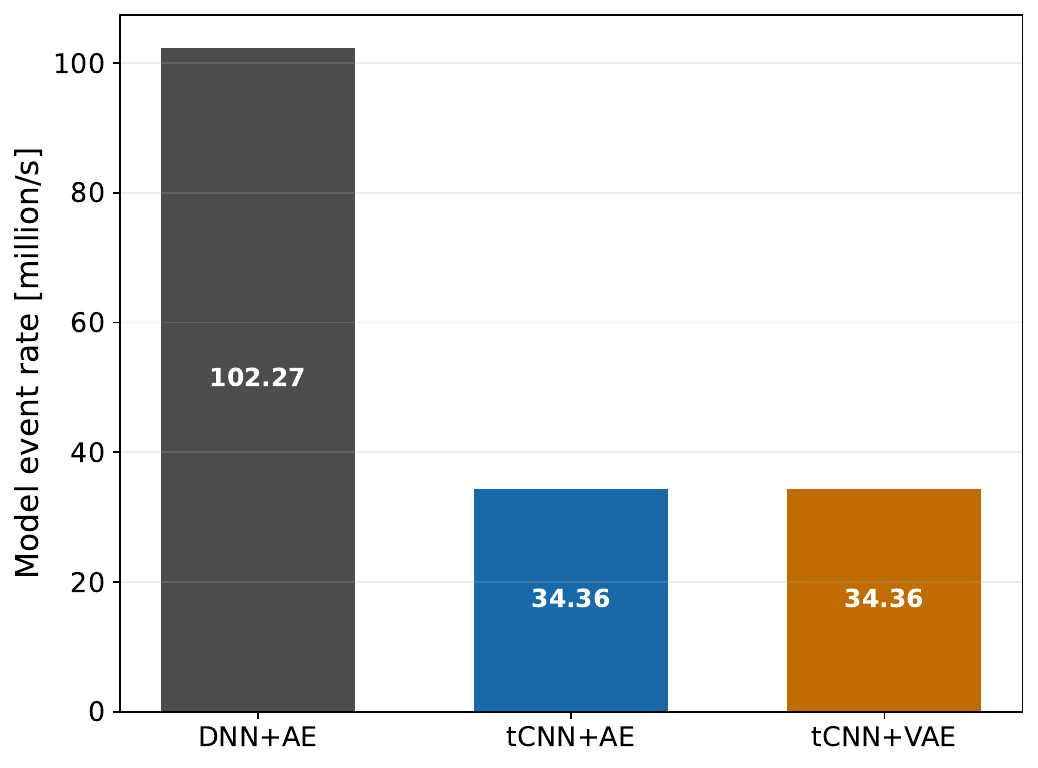}\\[-1mm]
        {\small (c) Model-input rate.}
    \end{minipage}\hfill
    \begin{minipage}[t]{0.49\textwidth}
        \centering
        \includegraphics[width=\linewidth]{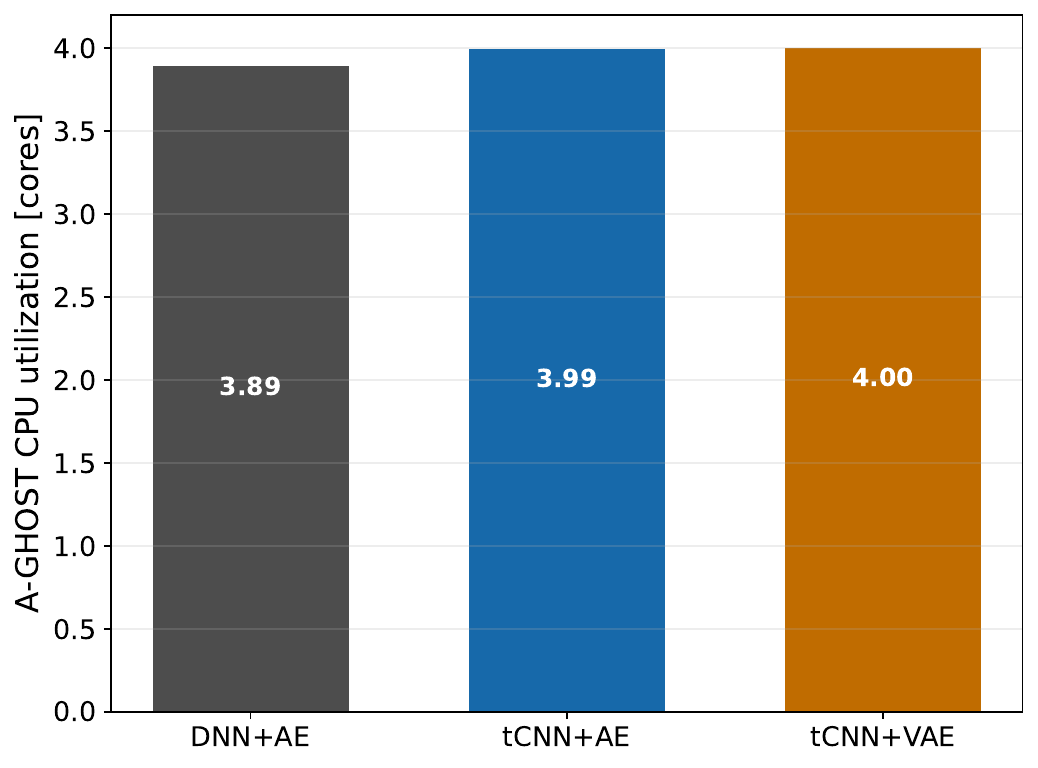}\\[-1mm]
        {\small (d) A-GHOST host CPU utilisation.}
    \end{minipage}
    \caption{Comparison of the three inference workloads with the streaming backend driven at the upper end of the rate scan.  The dense event-wise autoencoder is denoted DNN+AE.  (a) Delivered receive throughput (with enabled inference).  (b) Mean TensorRT inference latency at the selected operating point.  (c) Rate of model inputs presented to TensorRT.  One temporal-model input contains three consecutive source events, while one DNN+AE input contains one source event.  (d) Host CPU utilisation of the complete A-GHOST process.}
    \label{fig:modelcomparison}
\end{figure}

For this reason we are subjecting the three models (DNN+AE, tCNN+AE and tCNN+VAE) at 100 Gbps and compare in Figure~\ref{fig:modelcomparison} the delivered receive throughput, mean inference latency, model-input rate and host CPU utilisation. All three configurations sustain a receive rate close to 100 Gbps: 98.91 Gbps for the DNN+AE and 99.69 Gbps for both temporal models. The difference by 0.8Gbps between the models have been measured to be due to CPU scheduling arising from the rate controlled transmitter. This level of variability is specific to the current setup and does not reduce the argument of computational freedom provided by the GPU backend. When the FPGA data source will be deployed the control over the throughput becomes more explicit. The receive path therefore remains close to the tested link limit even though the downstream inference workload changes substantially. The batch size for each model was selected such tht its inference pipeline could sustain the requested 100 Gbps input rate.

The measured mean inference latency is 0.170 ms for the DNN+AE, 0.112 ms for the tCNN+AE and 0.118 ms for the tCNN+VAE. These values are operating-point measurements of the batched TensorRT pipelines and should not be read as an intrinsic ordering of model complexity. In particular, the models use different input structures and different batch sizes. The relevant system-level result is that all three pipelines remain in a sub-millisecond inference regime while the receive path is kept close to its maximum tested rate.

The model-input rates in Figure~\ref{fig:modelcomparison}(c) also reflect the different input definitions. The DNN+AE processes 102.27 million single-event inputs per second, whereas the tCNN+AE and tCNN+VAE each process 34.36 million three-event windows per second. Since each temporal window contains three consecutive source events, the latter corresponds to approximately 103.1 million source events per second before grouping. The apparent factor-of-three difference is therefore a consequence of the temporal model definition and not a factor-of-three reduction in the incoming physics event stream.

The host-side cost is nearly independent of the model choice in these measurements. The A-GHOST process uses 3.89 equivalent CPU cores with the DNN+AE, 3.99 cores with the tCNN+AE and 4.00 cores with the tCNN+VAE. This weak dependence, imposed by construction of DAQIRI, is consistent with the host retaining the same role across all three tests: coordinating DAQIRI reception, recording delivered spans and freeing the memory after inference on the GPU, while the model dependent computation is executed in the accelerator domain. It also indicates that the received 100 Gbps stream is completely bypassing the CPU of the host, allowing direct GPU inference on the received stream.

\subsection{Batch size and delivered network throughput}

\begin{figure}[h]
    \centering
    \includegraphics[width=0.85\textwidth]{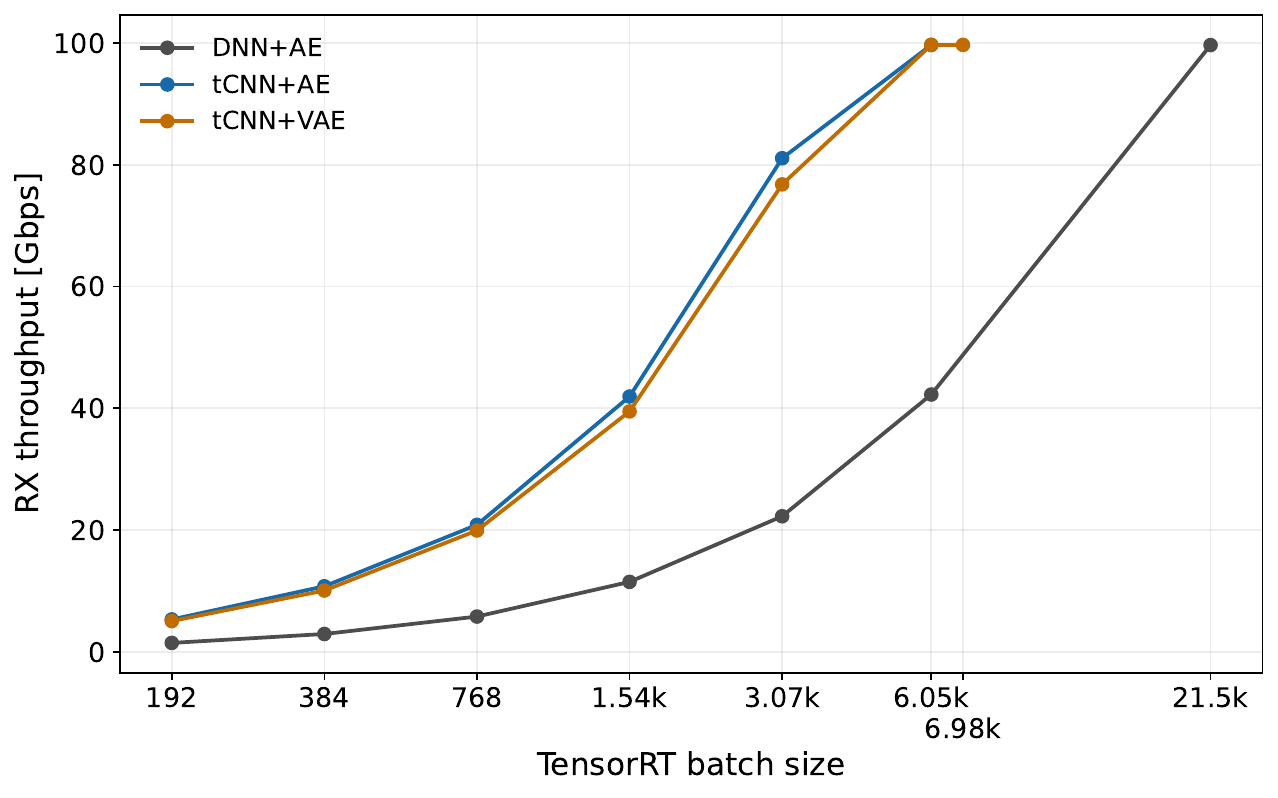}
    \caption{Delivered receive throughput versus TensorRT batch size, where a batch size is defined as the total number of inputs for a single inference of the models.  Temporal-model inputs contain three consecutive source events, while DNN+AE inputs contain one.}
    \label{fig:modelbatch}
\end{figure}

The comparison across models also makes the role of batching explicit. Figure~\ref{fig:modelbatch} shows that delivered receive throughput as a function of batch size. At small batches, launch, synchronisation and buffer-turnover overheads limit how quickly the application can consume the incoming stream. Increasing the batch size amortises these costs and raises the sustainable receive rate until a plateau near the link limit is reached.

At a batch size of approximately $3.1\times10^3$ model inputs, the temporal workloads already sustain about 76--81 Gbps, while the event-wise DNN+AE is near 22 Gbps. At approximately $6.1\times10^3$ model inputs per batch, the temporal pipelines reach the high-rate plateau close to 100 Gbps, whereas the DNN+AE is still near 42 Gbps and reaches the same regime only with a batch of order $2.2\times10^4$. A substantial part of this shift follows directly from the event representation. The temporal input carries three source events, so a batch containing the same number of model inputs represents roughly three times as much source-event payload.

Taken together, Figures~\ref{fig:modelcomparison} and~\ref{fig:modelbatch} show that the  networking and GPU-assembly path is not tied to a single inference architecture. The same backend can feed an event-wise dense autoencoder, a temporal autoencoder and a generative temporal model while maintaining high delivered network throughput once an appropriate software batching point is selected. This separation between the transport interface and the accelerator-side algorithm is central to the A-GHOST concept. A fixed-latency source can retain a stable output format while the downstream model, temporal context and batching strategy evolve in software.

\section{Discussion and outlook}

The proof of concept establishes the principal systems ingredients required by A-GHOST. First, a compact HEP-derived representation can be transported continuously at rates approaching 100 Gbps into an accelerator-capable node. Second, packet-oriented delivery can be transformed on the GPU into persistent inference windows so that network fragmentation does not dictate the model batch size. Third, the resulting stream can be processed by dense and temporal convolutional neural network architectures whose complexity and buffering requirements would be difficult to accommodate within the resource and latency envelope of the hardware trigger itself.

The distinction from a conventional hardware trigger is therefore one of role rather than replacement. FPGAs remain the appropriate technology for deterministic timing, fixed-latency preprocessing and operations that must complete before a hardware decision deadline. A-GHOST investigates what becomes possible once a compact representation is exported at line rate to a second, software-defined domain. The GPU stage can use more memory, can maintain context over multiple events, can trade batching latency against throughput, and can be updated without changing trigger firmware. This flexibility also opens the possibility of event-level data-quality monitoring and curation, using temporal context to identify transient detector effects and flag affected events for downstream analysis.

The present measurements stop at the backend boundary. The transmitter and receiver reside on the IGX Thor, connected through an external cable, so the next milestone is to replace the software transmitter with an FPGA transmitter and validate framing, synchronisation, backpressure behaviour and long-duration stability across the complete FPGA-to-GPU link. Further work should also measure the full source-to-inference latency distribution, optimise packet size and inference batch size jointly, evaluate several concurrent model streams, and determine how the architecture scales when several high-rate links are aggregated into one or more accelerators. In the scale up to multiple high-rate sources, latest features in high-performance networking can be applied to main high-throughput and low latency. For example, hardware-based packet reordering on the NIC can offload packet-handling tasks from the GPU, preserving more compute for inference.

In this sense, the result reported here is intended as a foundation for A-GHOST rather than a final trigger design. It demonstrates a viable proof-of-concept architecture in which HEP-derived data can be streamed out of the fixed-latency domain at high rate and immediately explored with models that are substantially more expressive than those normally assigned to that domain.

\acknowledgments

This work has been partially funded by the Eric \& Wendy Schmidt Fund for Strategic Innovation through the CERN Next Generation Triggers project under grant agreement number SIF-2023-004. We also gratefully acknowledge the support of the UK's Science and Technology Facilities Council (STFC). Students from UCL are supported by the STFC UCL Centre for Doctoral Training in Data Intensive Science  (ST/W00674X/1) and by UCL and industry funds. DWM is supported by the National Science Foundation under Grant No. PHY-2310094. Explicit mention to CERN Openlab for facilitating the collaboration between NVIDIA and CERN for the development of this work. In addition we acknowledge the contribution of the researcher Dr. Petr Manek at the Institute of Experimental \& Applied Physics at CTU Prague and the UCL project student Rafay Muhammand for their significant contribution during the early days of A-GHOST and specially for the development of the SPOOK monitoring container. 

\paragraph{Funding:} The contributors of this paper are supported by:
\begin{enumerate}
    \item Eric \& Wendy Schmidt Fund for Strategic Innovation (SIF-2023-004)
    \item Science and Technology Facilities Council (STFC) (ST/W00674X/1)
    \item National Science Foundation (PHY-2310094)
\end{enumerate}

\end{document}